\documentclass{article}
\usepackage{ijcai24}

\usepackage{times}
\usepackage{soul}
\usepackage{url}
\usepackage[hidelinks]{hyperref}
\usepackage[utf8]{inputenc}
\usepackage[small]{caption}
\usepackage{graphicx}
\usepackage{subfigure}
\usepackage{amsmath}
\usepackage{amsthm}
\usepackage{amssymb}
\usepackage{booktabs}
\usepackage{algorithm}
\usepackage{algorithmicx}
\usepackage{algpseudocode}
\usepackage[switch]{lineno}
\usepackage{amsfonts}
\usepackage{textcomp}
\usepackage{color}
\usepackage{multirow}
\usepackage{mathrsfs}
\usepackage{array}
\usepackage{threeparttable}
\usepackage{float}
\usepackage{algpseudocode}
\usepackage{eqparbox}
\usepackage{bm}
\usepackage{makecell}
\usepackage{diagbox}
\usepackage{stfloats}
\usepackage{url}
\usepackage{verbatim}
\usepackage{setspace}

\begin{document}
	\title{Enhancing Cross-Dataset EEG Emotion Recognition: A Novel Approach with Emotional EEG Style Transfer Network}
	\author{
		Yijin Zhou$^1$
		\and
		Fu Li$^1$\and
		Yang Li$^{1,*}$\and
		Youshuo Ji$^1$\and
		Lijian Zhang$^2$\And
		Yuanfang Chen$^2$\\
		\affiliations
		$^1$Department of Artificial Intelligence, Xidian University\\
		$^2$Beijing Institute of Mechanical Equipment\\
		\emails
		zhouyijin@stu.xidian.edu.cn, liy@xidian.edu.cn
	}
	\maketitle
	\begin{abstract}
		Recognizing the pivotal role of EEG emotion recognition in the development of affective Brain-Computer Interfaces (aBCIs), considerable research efforts have been dedicated to this field. While prior methods have demonstrated success in intra-subject EEG emotion recognition, a critical challenge persists in addressing the style mismatch between EEG signals from the source domain (training data) and the target domain (test data). To tackle the significant inter-domain differences in cross-dataset EEG emotion recognition, this paper introduces an innovative solution known as the Emotional EEG Style Transfer Network (E$^2$STN). The primary objective of this network is to effectively capture content information from the source domain and the style characteristics from the target domain, enabling the reconstruction of stylized EEG emotion representations. These representations prove highly beneficial in enhancing cross-dataset discriminative prediction. Concretely, E$^2$STN consists of three key modules\textemdash transfer module, transfer evaluation module, and discriminative prediction module\textemdash which address the domain style transfer, transfer quality evaluation, and discriminative prediction, respectively. Extensive experiments demonstrate that E$^2$STN achieves state-of-the-art performance in cross-dataset EEG emotion recognition tasks.
	\end{abstract}

	\section{Introduction}
	
	In the 21st century, brain-computer interface (BCI) technology emerges as a novel avenue for human-computer interaction, offering a novel communication paradigm against the background of the burgeoning metaverse~\cite{guo2022metaverse}. Given the pivotal role of emotion in human-computer interaction, affective Brain-Computer Interfaces (aBCIs) have attracted significant attention across interdisciplinary fields~\cite{FIORINI2020105217}. The aBCIs predominantly rely on two modalities\textemdash behavioral signals and physiological signals\textemdash for emotion recognition~\cite{he2020advances}. Compared with behavioral signals, such as facial expressions, speech, and text, it is more reliable to distinguish the spontaneous emotion state through physiological signals, such as electrocardiogram (ECG), electrooculogram (EOG), electromyogram (EMG), and electroencephalogram (EEG)~\cite{song2021variational}. Among these physiological signals, EEG signals originating in the cerebral cortex are particularly associated with spontaneous emotional states~\cite{he2020advances}. And with the development of wearable non-invasive EEG acquisition equipment in recent years, more and more researches are focusing on the field of EEG emotion recognition.
	
	\begin{figure}[htp]
		\centering
		{\includegraphics[width=1\linewidth]{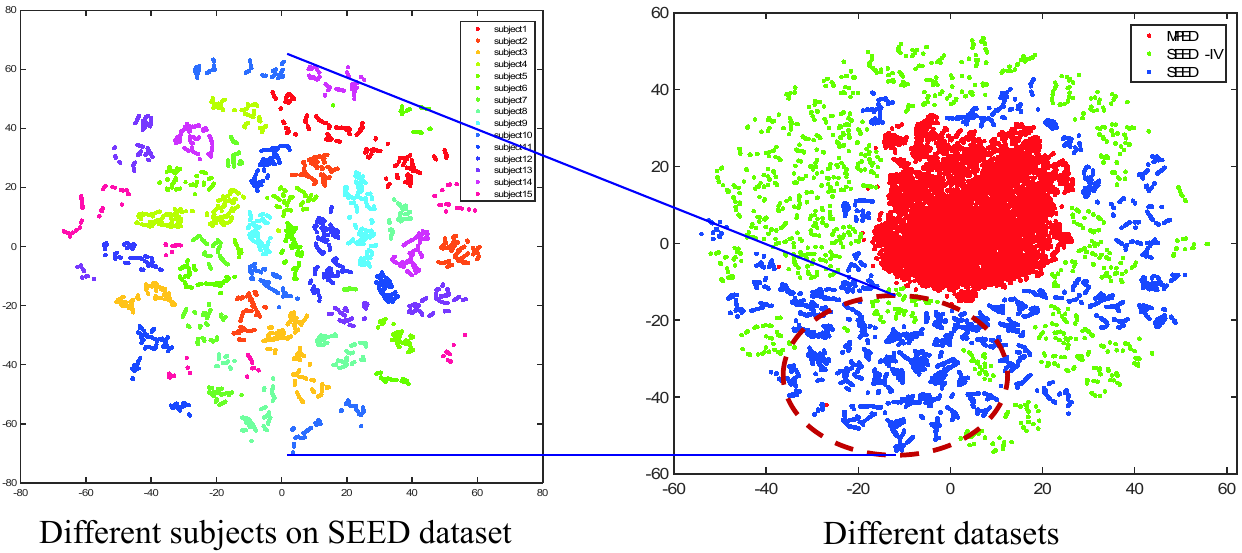}}
		\caption{\label{TSNE_domain_shift} Distribution of EEG data in different subjects and datasets.}
	\end{figure}
	
	The existing research on EEG emotion recognition has predominantly concentrated on intra-subject tasks~\cite{xiao20224d}. For instance, considering the abundant saptial information in EEG signals, Song et al. proposed to convert multi-channel EEG signals into an image format, which converts the question of EEG emotion recognition into image recognition. In this regard, they introduced a novel EEG-to-image method and a graph-embedded convolutional neural network (GECNN) approach. The effectiveness of GECNN was validated through extensive experiments on four public datasets~\cite{9448460}. Additionally, Zhou et al. introduced a Progressive Graph Convolution Network (PGCN) for EEG emotion recognition, leveraging insights from neuroscience on dynamic brain relationships. PGCN achieves state-of-the-art performance by progressively learning discriminative features from coarse- to fine-grained emotion categories~\cite{ZHOU2023126262}. Despite the numerous methods proposed for EEG emotion recognition in recent years, there are significant issues that merit thorough investigation to advance this field. The primary concern is the protocol for EEG emotion recognition. Existing protocols for EEG emotion recognition mostly involve intra-subject and cross-subject classification tasks, where training and test EEG data originate from the same experimental environment. The performance variation across different experimental environments, especially in cross-dataset EEG emotion recognition, needs further exploration. To clearly and intuitively show the differences in EEG data distribution, T-SNE technology was employed to visualize the EEG data of different subjects in different datasets, as illustrated in Fig.~\ref{TSNE_domain_shift}. Notably, substantial differences exist in the EEG data distribution among different subjects of the same dataset, which are more significant among diverse datasets.
	
	The second critical issue involves addressing domain differences. Recent studies have attempted to tackle domain shift in cross-subject EEG emotion recognition tasks. For example, Li et al. proposed a multisource transfer learning method, treating existing subjects as sources and the new subject as a target to achieve style transfer mapping~\cite{8675478}. Advanced performance in addressing distribution differences between training and test data in cross-subject EEG emotion recognition tasks has been demonstrated by methods like BiDANN~\cite{li2018novel} and TANN~\cite{LI202192}. However, the inter-domain differences in cross-dataset EEG emotion recognition surpass those observed in the cross-subject EEG emotion recognition task, as depicted in Fig.~\ref{TSNE_domain_shift}. Minimizing these differences between domains holds promise for improving cross-dataset EEG emotion recognition and enhancing generalization to new emotional EEG data.
	
	To tackle these issues, we propose an Emotional EEG Style Transfer Network (E$^2$STN) in this study to obtain stylized emotional EEG representations. These representations encapsulate emotion content of the source domain and style characteristics of the target domain, enabling the model to make discriminative predictions for cross-dataset emotional EEG samples. Specifically, E$^2$STN comprises three unique modules: the transfer module, transfer evaluation module, and discriminative prediction module. The transfer module reorganizes emotional pattern information from the source domain and statistical style from the target domain to generate new stylized EEG representations. The transfer evaluation module, incorporating content-aware loss, style-aware loss, and identity loss, ensures the precise fusion of information from both domains. The discriminative prediction module, utilizing a dynamic graph convolutional network and fully connected layers, extracts deep features for discriminative predictions. Joint optimization with cross-entropy loss and transfer evaluation losses guides the entire model for comprehensive cross-dataset EEG emotion recognition.
	
	To the best of our knowledge, this work is the first to reorganize emotional content information from the source domain and	statistical characteristic style from the target domain into new stylized EEG representations to enhance cross-dataset EEG emotion recognition. The proposed E$^2$STN generates stylized emotional EEG representations and further performs discriminative predictions from source domain and stylized representations. The joint loss optimization ensures precise fusion of complementary information and guides discriminative prediction for cross-dataset EEG emotion recognition. Extensive experiments validate E$^2$STN's state-of-the-art performance in cross-dataset EEG emotion recognition tasks.
	
	\section{Proposed Method for Emotion Recognition}
	\label{Sec: The proposed method}
	
	\begin{figure*}[h]
		\centering {\includegraphics[width=1\linewidth]{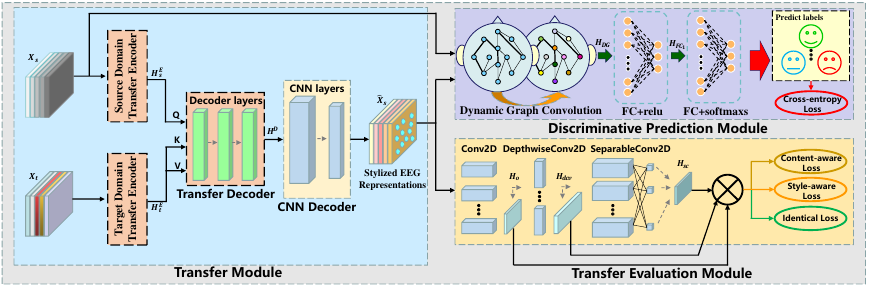}}
		\caption{\label{EESTN framework} Framework of E$^2$STN. E$^2$STN consists of three modules to obtain stylized emotional EEG representations containing emotion content of the source domain and statistical characteristics of the target domain, and meanwhile, make discriminative predictions for cross-dataset emotional EEG samples.}
	\end{figure*}
	
	To elucidate the proposed method, we present the framework of E$^2$STN in Fig.~\ref{EESTN framework}. The primary objective of this network is to restructure the emotional content information from the source domain and the statistical characteristic style from the target domain, yielding newly stylized source domain EEG representations. These representations are crucial for the successful execution of cross-dataset EEG emotion recognition tasks. We adopt three key modules to achieve this goal, i.e., the transfer module, transfer evaluation module, and discriminative prediction module. The transfer module focuses on generating stylized emotional EEG representations that encapsulate both the emotional content information from the source domain and the statistical characteristic style from the target domain. Subsequently, the discriminative prediction module processes the source domain and stylized EEG representations for effective cross-dataset EEG emotion recognition. Simultaneously, the transfer evaluation module extracts multi-scale spatio-temporal features from the source domain and stylized EEG representations, constructing multi-dimensional losses that intricately guide the emotional EEG style transfer process. In the following, we introduce the details of the proposed E$^2$STN model.	 
	
	\subsection{Obtaining Stylized Emotional EEG Samples}
	
	To obtain emotional EEG representations that contain both the emotional content information of the source domain and the statistical characteristics style of the target domain, the transfer process is divided into two essential steps. The initial step involves constructing transfer encoders corresponding to the source and target domains. The two encoders capture the global dependencies within the domain-specific information of distinct fields (i.e., the emotional content of the source domain and the style characteristics of the target domain), respectively. Drawing inspiration from the Transformer method~\cite{NIPS2017_3f5ee243}, the encoders of E$^2$STN employ multi-head self-attention layers to assign dynamic weights to different EEG channels. This dynamic weighting, guided by domain-specific information, highlights the more significant electrode dependencies within the specific domain. Such dynamic dependencies, enriched with domain-specific information, have stronger capabilities in representing their corresponding domain characteristics. The second crucial step in the transfer process is the fusion of source domain content information and target domain style information within the decoder, yielding stylized emotional EEG features. The decoder achieves this by iteratively combining content and style information through a multi-layer structure, applying the target domain style to the source domain EEG features. The use of the residual connection method in the decoder layer ensures that the content information of the source domain remains undistorted throughout the fusion process. Finally, a CNN decoder, comprising multiple convolutional layers, reconstructs the stylized EEG features into representations of the same dimension as the input.
		
\begin{figure}[h]
		\centering 
		\subfigure[Encoder layer]{\includegraphics[width=0.8\linewidth]{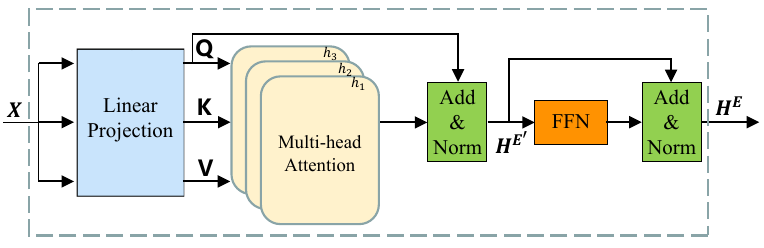}}
		\quad
		\subfigure[Decoder layer]{\includegraphics[width=1\linewidth]{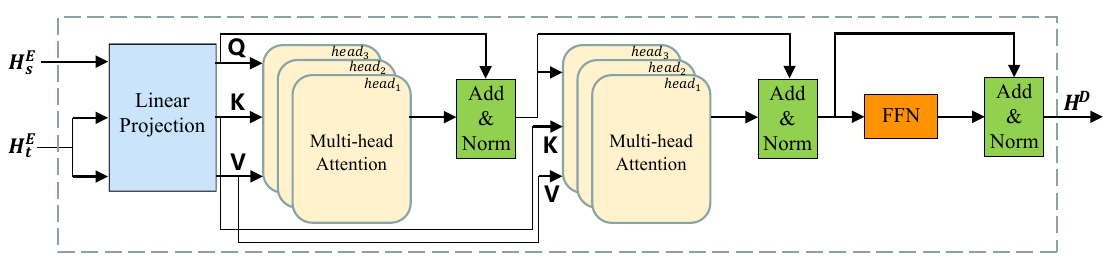}}
		\caption{\label{encoder framework} Architecture of the encoder and decoder layer in the transfer module. "Add\&Norm" signifies the addition of the residual connection and subsequent layer normalization.}
	\end{figure}

	Specifically, we utilize the $B$ frequency bands EEG representations after pre-decomposition of the raw EEG signals. The input emotional EEG representations of the proposed model corresponding to the source and target domains are denoted as $ {\rm \mathbf{X_s}} \in \mathbb{R} ^ {C\times B} $ and $ {\rm \mathbf{X_t}} \in \mathbb{R} ^ {C\times B} $, respectively, where $C$ is the number of EEG channels. Two corresponding encoders are employed to extract domain-specific information from their respective source and target domain EEG representations. The structure of the encoder layer is illustrated in Fig.~\ref{encoder framework} (a). $ {\rm \mathbf{X_s}} $ and $ {\rm \mathbf{X_t}} $ undergo encoding into query (${\rm \mathbf{Q}}$), key (${\rm \mathbf{K}}$), and value (${\rm \mathbf{V}}$) vectors within their respective encoders. The subsequent explanation focuses on $ {\rm \mathbf{X_s}} $ to illustrate the encoding process, as shown in fomula~(\ref{encoder 1}),~(\ref{encoder 2}),~(\ref{encoder 3}),~(\ref{encoder 4}).
	
	\begin{equation}
		{\rm \mathbf{Q_s^E}} = {\rm \mathbf{X_s}}{\rm \mathbf{W_s^q}}, {\rm \mathbf{K_s^E}} = {\rm \mathbf{X_s}}{\rm \mathbf{W_s^k}}, {\rm \mathbf{V_s^E}} = {\rm \mathbf{X_s}}{\rm \mathbf{W_s^v}},
		\label{encoder 1}
	\end{equation}
	where $ {\rm \mathbf{W_s^q}}, {\rm \mathbf{W_s^k}}, {\rm \mathbf{W_s^v}} \in \mathbb{R} ^ {B\times m} $ are trainable linear projection matrices. To enable the encoder to pay attention to the information from different channels, $ {\rm \mathbf{Q_s^E}} $, $ {\rm \mathbf{K_s^E}} $, and $ {\rm \mathbf{V_s^E}} $ vectors are divided into several attention heads, that $ {\rm \mathbf{Q_s^E}}, {\rm \mathbf{K_s^E}}, {\rm \mathbf{V_s^E}} \in \mathbb{R} ^ {h \times C \times p} $, where $ h = \frac{m}{p} $ is the number of attention heads. Then the multi-head self-attention (MSA) can be calculated by:
	
	\begin{equation}
		{\rm \mathbf{M_s^E}} = [h_1, \cdots, h_h]{\rm \mathbf{W_s^O}} \in \mathbb{R} ^ {C\times m},
		\label{encoder 2}
	\end{equation}
	where $ [h_1, \cdots, h_h] $ represents the output of each attention head.
	
	To maintain domain-specific information, the MSA matrix is added to the Q vector and subsequently subjected to layer normalization. This operation can be expressed as:
	
	\begin{equation}
		{\rm \mathbf{H_s^{E^{\prime}}}} = LayerNorm({\rm \mathbf{M_s^E}}({\rm \mathbf{Q_s^E}}, {\rm \mathbf{K_s^E}}, {\rm \mathbf{V_s^E}}) + {\rm \mathbf{Q_s^E}}),
		\label{encoder 3}
	\end{equation}
	
	\begin{equation}
		{\rm \mathbf{H_s^E}} = LayerNorm(FFN({\rm \mathbf{H_s^{E^{\prime}}}}) + {\rm \mathbf{H_s^{E^{\prime}}}}) \in \mathbb{R} ^ {C \times m},
		\label{encoder 4}
	\end{equation}
	where $ FFN(\cdot) $ is a fully connected feed-forward network. Similarly, we can easily obtain the domain-specific features of the target domain $ {\rm \mathbf{H_t^E}} $ through the above formulas.
	
	To integrate the emotional content information of the source domain and the statistical characteristic style of the target domain, we devise a three-layer transfer decoder, which applies the style of the target domain to the emotional features of the source domain progressively. The structure of a single decoder layer is depicted in Fig.~\ref{encoder framework} (b). The source domain features $ {\rm \mathbf{H_s^E}} $, encapsulating the emotional content information of the source domain, serve as the primary focus of transfer. These features are utilized as the query vectors for the first decoder layer. To make the source domain features more similar to the taget domain style, the target domain features $ {\rm \mathbf{H_t^E}} $ are employed as key and value vectors for the first decoder layer, which calculates a similarity matrix with the query vectors to weigh the emotional content features $ {\rm \mathbf{H_s^E}} $. Specifically, $ {\rm \mathbf{Q_1^D}} $, $ {\rm \mathbf{K_1^D}} $, and $ {\rm \mathbf{V_1^D}} $ are obtained through linear projection, as illustrated in formula~(\ref{decoder 1}).
	
	\begin{equation}
		{\rm \mathbf{Q_1^D}} = {\rm \mathbf{H_s^E}}{\rm \mathbf{W_q^D}}, {\rm \mathbf{K_1^D}} = {\rm \mathbf{H_t^E}}{\rm \mathbf{W_k^D}}, {\rm \mathbf{V_1^D}} = {\rm \mathbf{H_t^E}}{\rm \mathbf{W_v^D}}.
		\label{decoder 1}
	\end{equation}
	Subsequently, two MSA layers and one fully connected feed-forward network (FFN) are employed in the first decoder layer with residual connections. The output of the first decoder layer is then passed on to the second decoder layer, and so forth. Consequently, we can readily derive the output $ {\rm \mathbf{H^D}} \in \mathbb{R} ^ {C \times m} $ of the transfer decoder using formulas~(\ref{decoder 1}),~(\ref{encoder 3}),~(\ref{encoder 4}).
	
	To restore the dimension of the stylized features, a two-layer Convolutional Neural Network (CNN) decoder is employed to refine the output of the transfer decoder $ {\rm \mathbf{H^D}} $. This process allows reshaping of the stylized EEG features $ {\rm \mathbf{H^D}} \in \mathbb{R} ^ {C \times m} $ into generated stylized EEG representations $ {\rm \mathbf{\hat{X}_s}} \in \mathbb{R} ^ {C \times B} $. These stylized EEG representations $ {\rm \mathbf{\hat{X}_s}} $ have the same emotion labels as their corresponding source domain EEG representations.
	
	\subsection{Obtaining Discriminative Features and Predictions}
	
	After obtaining the stylized source-domain EEG representations, a dynamic graph network is constructed to extract deep features, enabling E$^2$STN to learn discriminative features from the source domain and stylized EEG representations. Both the source EEG representations $ {\rm \mathbf{X_s}} $ and the corresponding stylized EEG representations $ {\rm \mathbf{\hat{X}_s}} $ are jointly input into the discriminative prediction module to obtain discriminative features and predictions. Concretely, the data-driven graph of the dynamic graph network is characterized by an adjacency matrix $ {\rm \mathbf{G}} $, which dynamically adjusts based on the input representations $ {\rm \mathbf{X_s}} $ and $ {\rm \mathbf{\hat{X}_s}} $. To ensure that $ {\rm \mathbf{G}} $ contains intra-channel spatial information and frequency band information, two trainable matrices $ {\rm \mathbf{W_s}} \in \mathbb{R} ^ {C \times C} $ and $ {\rm \mathbf{W_f}} \in \mathbb{R} ^ {B \times (C*B)} $ are left- and right-multiplied by the input features, respectively. This process can be expressed as follows:
	
	\begin{equation}
		{\rm \mathbf{G}} = ReLU[({\rm \mathbf{W_s}}{\rm \mathbf{X}} + {\rm \mathbf{B}}){\rm \mathbf{W_f}}] \in \mathbb{R} ^ {C \times (C*B)},
		\label{G_d}
	\end{equation}
	where $ {\rm \mathbf{X}} = {\rm \mathbf{X_s}}/{\rm \mathbf{\hat{X}_s}} $, $ ReLU $ is applied to the output to guarantee non-negative elements, the bias matrix $ {\rm \mathbf{B}} \in \mathbb{R} ^ {C \times B} $ is used to increase the flexibility of graph structure representation. Then, the adjacency matrix  $ {\rm \mathbf{G}} $ is reshaped into $ B $ adjacency matrices, i.e., $ {\rm \mathbf{G}} = [{\rm \mathbf{G}_{1}^{*}}, \dots, {\rm \mathbf{G}_{B}^{*}}] \in \mathbb{R} ^ {C \times C \times B} $, to represent graphs in $ B $ frequency bands.
	
	To avoid the high computational complexity associated with direct graph Fourier transform based on graph filtering theories, we adopt Chebyshev polynomials to approximate the graph convolution operation~\cite{kipf2017semisupervised}. Let $ \varphi_{k}(\mathbf{G}) = \mathbf{G}^{k} $ denotes the $ k $-order polynomial of the adjacency matrix $ \textbf{G} $. Consequently, the high-level features extracted by the dynamic Graph Convolutional Network (GCN) can be expressed as follows:
	
	\begin{equation}
		{\rm \mathbf{H_{DG}}} = \mathcal{G} * \mathcal{F} = \sum_{k=0}^{K-1}\varphi_{k}({\rm \mathbf{G}}){\rm \mathbf{X}} \in  \mathbb{R} ^ {C \times F},
		\label{GCN_op}
	\end{equation}
	where $ \varphi_{k}({\rm \mathbf{G}}) $ is the $ k $-th level graph, $ F $ is the output dimension for the graph convolution operation. 
	
	Following that, the discriminative prediction module, acting as the supervision term of the E$^2$STN, employs two fully connected (FC) layers to predict the class labels. Therefore, the output of the second FC layer $ {\rm \mathbf{H_{FC}}} \in  \mathbb{R} ^ {1 \times P} $ can be easily deduced, where $ P $ is the output dimension of the FC layer. Consequently, the cross-entropy loss of E$^2$STN, aiming to achieve cross-dataset EEG emotion recognition, can be expressed as:
	
	\begin{equation}
		L_{ce} = -\sum_{i}^{P}y_i log{\hat{y_i}}
		\label{coarse_loss},
	\end{equation}
	where $ {\rm \mathbf{Y}} = \{y_1, \dots, y_P\} \in \mathbb{R} ^ {1 \times P} $ represents the ground-truth label; $ {\rm \mathbf{\hat{Y}}} = \{\hat{y_1}, \dots, \hat{y}_P\} \in \mathbb{R} ^ {1 \times P} $ is the discriminative prediction from the softmax layer of E$^2$STN.
	
	\subsection{Multi-objective Joint Optimization}
	
	To optimize stylized emotional EEG representations, we specially propose a transfer evaluation module to constrain the style transfer process. In the emotional style transfer process, we primarily consider three factors, leading to the formulation of three corresponding losses: content-aware loss $ L_c $, style-aware loss $ L_s $, and identity loss $ L_{id} $.	The first crucial consideration is preserving the emotional content information of the source domain during the transfer process. We regard the features extracted by the convolutional layer as containing the content information of the respective domain~\cite{Gatys_2016_CVPR}. Therefore, the content-aware loss is constructed from the features extracted by multiple unique convolutional layers in the transfer evaluation module. This loss can be expressed as:
	\begin{equation}
		L_c = \frac{1}{n}\sum_{i=1}^{n}\left\| f_i({\rm \mathbf{\hat{X}_s}}) - f_i({\rm \mathbf{X_s}}) \right\|_2,
		\label{loss content}
	\end{equation}
	where $ f_i(\cdot) $ denotes the convolution operation function of the $ i $-th layer in the transfer evaluation module, and $ \left\|\cdot\right\|_2 $ represents the $ \ell_2 $-norm.
	
	Another crucial aspect in the transfer process is ensuring that the style characteristics of the stylized emotional EEG representations closely resemble those of the target domain. The Gram matrix of features extracted by the convolutional layer is considered representative of the statistical characteristics of the target domain~\cite{Gatys_2016_CVPR}. Consequently, we construct the style-aware loss for the target domain by considering the statistics (e.g., mean and variance) of each convolutional layer in the transfer evaluation module.
	\begin{equation}
		\begin{split}
			L_s = \frac{1}{n}\sum_{i=1}^{n}(\left\| \mu(f_i({\rm \mathbf{\hat{X}_s}})) - \mu(f_i({\rm \mathbf{X_t}})) \right\|_2 + \\
			\left\| \sigma(f_i({\rm \mathbf{\hat{X}_s}})) - \sigma(f_i({\rm \mathbf{X_t}})) \right\|_2),
		\end{split}
		\label{loss style}
	\end{equation}
	where $ \mu(\cdot) $ and $ \sigma(\cdot) $ denote the mean and variance of the features, respectively.
	
	In the final consideration, aiming to preserve more accurate content and style information in self-style transfer, we introduce an identity loss to ensure the undistorted nature of stylized EEG representations during the progressive transfer process. Specifically, for lossless and unbiased transfer of emotional EEG representations, we input the same representation $ {\rm \mathbf{X_s}} / {\rm \mathbf{X_t}} $ into the source and target domain transfer encoders. The resulting stylized emotional EEG representation, denoted as $ {\rm \mathbf{\hat{X}_{ss}}} / {\rm \mathbf{\hat{X}_{tt}}} $, should be identical to the original $ {\rm \mathbf{X_s}} / {\rm \mathbf{X_t}} $. Consequently, the identity loss $ L_{id} $ can be defined as:
	\begin{equation}
		\begin{split}
			L_{id} = \frac{1}{n}\sum_{i=1}^{n} (\left\| f_i({\rm \mathbf{\hat{X}_{ss}}}) - f_i({\rm \mathbf{X_s}}) \right\|_2 + \left\| f_i({\rm \mathbf{\hat{X}_{tt}}}) - f_i({\rm \mathbf{X_t}}) \right\|_2),
		\end{split}
		\label{loss id}
	\end{equation}
	
	To ensure that the features extracted by multiple convolutional layers in the transfer evaluation module contain multi-dimensional and multi-scale spatio-temporal information, we employ three distinct convolution convolutional kernels to construct the convolutional network.
	
	\begin{enumerate}
		\item \textbf{2D Convolutional Layer:} The first layer utilize a convolution kernel of size $ (1,3) $, exploring latent relationships between key bands of stylized EEG features. The output is represented as $ {\rm \mathbf{H_{c}}} \in \mathbb{R}^{C \times B \times F_1} $, with $ F_1 $ being the number of convolutional filters.
		\item \textbf{Depthwise Convolution:} This layer employs a convolutional kernel of size $ (C, 1) $ to capture the spatial information of stylized EEG features. It generates depth features $ {\rm \mathbf{H_{dc}}} \in \mathbb{R}^{1 \times B \times (F_1*D)} $, where $ D $ is a depth parameter controlling the number of spatial filters.
		\item \textbf{Separable convolution:} Building upon depthwise convolution, separable convolution utilizes a convolution kernel of size $ (1,3) $ and $ F_2 $ pointwise convolutions to optimally merge the spatial features. This process compresses the feature $ {\rm \mathbf{H_{dc}}} $ into $ {\rm \mathbf{H_{sc}}} \in \mathbb{R}^{1 \times B \times F_2} $ in the channel dimension.
	\end{enumerate}
	where $ {\rm \mathbf{H_{c}}} $, $ {\rm \mathbf{H_{dc}}} $, and $ {\rm \mathbf{H_{sc}}} $ correspond to the features extracted by each convolutional layers in formulas~(\ref{loss content}),~(\ref{loss style}), and~(\ref{loss id}), respectively.
	
	In conclusion, the transfer losses in the transfer evaluation module and the cross-entropy loss in the discriminative prediction module collectively form a multi-objective joint optimization loss function $ L $.
	
	\begin{equation}
		L = L_c + \lambda L_s + \nu L_{id} + \xi L_{ce},
	\end{equation}
	where $ \lambda, \nu, \xi $ are hyper-parameters used to control the proportion between the optimization loss functions. E$^2$STN is iteratively optimized by minimizing $ L $, and the emphasis on transferring tasks and classification tasks is achieved by adjusting the hyperparameters. The training procedure for E$^2$STN is outlined in the Algorithm 1 of Appendix A. Further implementation details of E$^2$STN are provided in Appendix C.
	
	\section{Experiments}
	\label{Sec: Experiment}
	
	\subsection{Experiment protocol}
	\label{Experiment setting}
	
	The objective of this paper is to investigate cross-dataset EEG emotion recognition tasks. In alignment with the principles of previous experiments, we establish groups of experiment protocols for cross-dataset EEG emotion recognition based on SEED~\cite{zheng2015investigating}, SEED-IV~\cite{8283814}, and MPED~\cite{song2019mped} datasets, with detailed information provided in Appendix B. Table I of Appendix B summarizes the setup details of the cross-dataset EEG emotion recognition experiments. In brief, to maintain category balance in the training samples, we choose neutral, sad, and happy (joy) emotions of SEED, SEED-IV, and MPED datasets for 3-category cross-dataset experiments. For 4-category cross-dataset experiments, we choose neutral, happy (joy), sad, and fear emotions from SEED-IV and MPED datasets. All subjects' samples from one dataset are considered as source domain data, while one subject's samples from another dataset are utilized as target domain data. This approach allows us to conduct experiments on two types of cross-dataset EEG emotion recognition tasks, involving three and four categories of emotions. For instance, we denote the 3-category cross-dataset EEG emotion recognition experiment using the MPED$^3$ dataset as the source domain and the SEED$^3$ dataset as the target domain as \emph{MPED$^3$ $\rightarrow$ SEED$^3$} in this paper.
	
	\subsection{Experiment Results}
	
	\subsubsection{3-category cross-dataset EEG emotion recognition}
	\label{3-category Cross-dataset EEG emotion recognition}
	To evaluate the performance of our model in cross-dataset EEG emotion recognition, we conduct extensive experiments following the specified protocols. In comparison with other advanced methods of EEG emotion recognition, we replicate the same experiments using 7 alternative methods. We either quote or reproduce their results from the literature to ensure a convincing comparison with the proposed method. The evaluation criteria for all subjects in the test dataset include mean accuracy (ACC) and standard deviation (STD). The experiment results are presented in Table~\ref{Table: cross-dataset}.
	
	\begin{table*}[h]
		\centering
		\fontsize{10}{14}\selectfont    %{字体尺寸}{行距}
		\caption{3-category cross-dataset classification performance for EEG emotion recognition on SEED, SEED-IV, and MPED.}
		\resizebox{\textwidth}{17mm}
		{
			\begin{tabular}{c|c|c|c|c|c|c}
				\hline
				\hline
				{\multirow{2}{*}{Method}}  &\multicolumn{6}{c}{\textbf{ACC / STD (\%)}} \cr
				
				\cline{2-7}
				&{MPED$^3$ $\rightarrow$ SEED$^3$} &{SEED-IV$^3$ $\rightarrow$ SEED$^3$} &{MPED$^3$ $\rightarrow$ SEED-IV$^3$} &{SEED$^3$ $\rightarrow$ SEED-IV$^3$} &{SEED-IV$^3$ $\rightarrow$ MPED$^3$} &{SEED$^3$ $\rightarrow$ MPED$^3$} \cr
				
				\hline	
				{SVM~\cite{suykens1999least}} &~48.94/04.96 &~23.63/10.28 &~27.71/02.93 &~29.47/06.27 &~33.32/0.08 &~40.53/04.74 \cr
				{BiDANN~\cite{li2018novel}} &~61.30/09.14 &~49.24/10.49 &~57.57/07.60 &~60.46/11.17 &~40.16/04.29 &~43.17/04.72 \cr
				{A-LSTM~\cite{song2019mped}} &~47.55/07.46 &~46.47/08.30 &~42.59/06.08 &~58.19/13.73 &~38.51/03.94 &~43.80/05.45 \cr
				{IAG~\cite{2020Instance}} &~60.89*/\textendash &~52.84/07.71 &~58.61/08.28 &~59.87/11.16 &~39.67/03.13 &~40.90*/\textendash \cr
				{TANN~\cite{LI202192}} &~64.23/09.63 &~58.41/07.16 &~55.14/09.59 &~60.75/10.61 &~37.16/01.69 &~40.62/04.66 \cr
				{GECNN~\cite{9448460}} &~62.90/06.58 &~58.02/07.03 &~60.88/06.96 &~57.25/07.53 &~38.82/03.52 &~43.15/03.08 \cr
				{PGCN-c~\cite{ZHOU2023126262}} &~63.02/09.37 &~58.45/07.37 &~56.97/07.89 &~60.87/13.20 &~39.95/05.14 &~43.27/04.99 \cr				
				{E$^2$STN} &~\textbf{73.51/07.23} &~\textbf{60.51/05.41} &~\textbf{62.32/06.60} &~\textbf{61.24/15.14} &~\textbf{40.43/04.49} &~\textbf{45.56/04.78} \cr
				\hline
				\hline
			\end{tabular}
		}
		\begin{tablenotes}
			\scriptsize
			%\centering
			{\item~~~ * indicates the results are obtained from the literature. The rest are obtained by our own implementation. The best result for each row in the Table is highlighted in boldface.}
		\end{tablenotes}
		\label{Table: cross-dataset}
	\end{table*}
	
	\begin{figure}[h]
		\centering
		\subfigure[\tiny MPED$^3$ $\rightarrow$ SEED$^3$]{\includegraphics[width=0.32\linewidth]{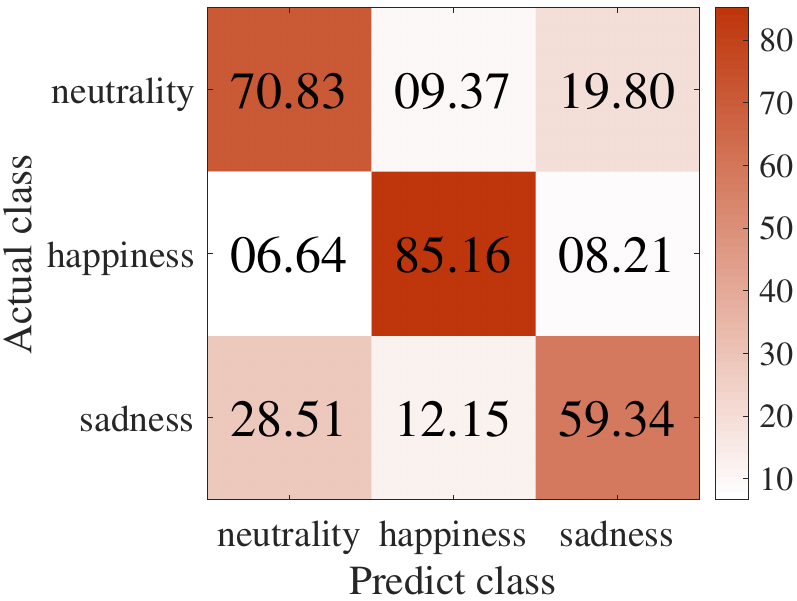}}
		\subfigure[\tiny SEED-IV$^3$ $\rightarrow$ SEED$^3$]{\includegraphics[width=0.32\linewidth]{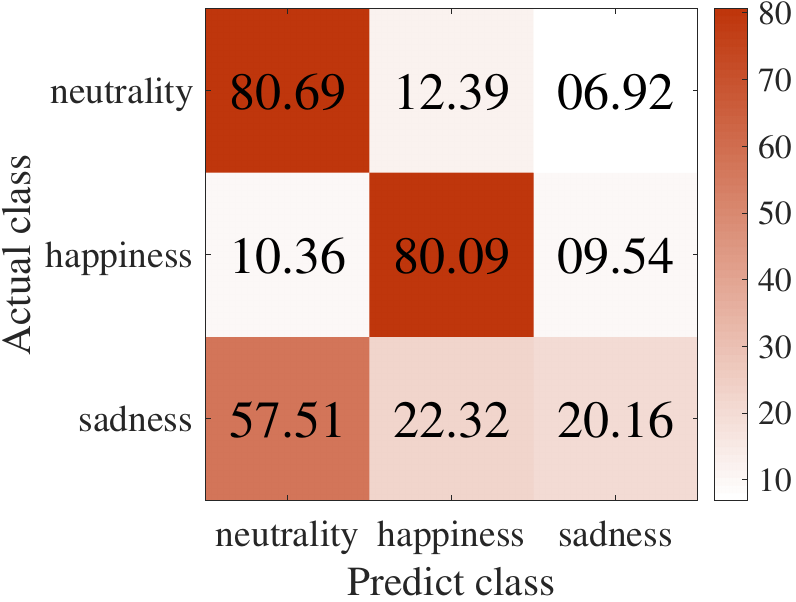}}
		\subfigure[\tiny MPED$^3$ $\rightarrow$ SEED-IV$^3$]{\includegraphics[width=0.32\linewidth]{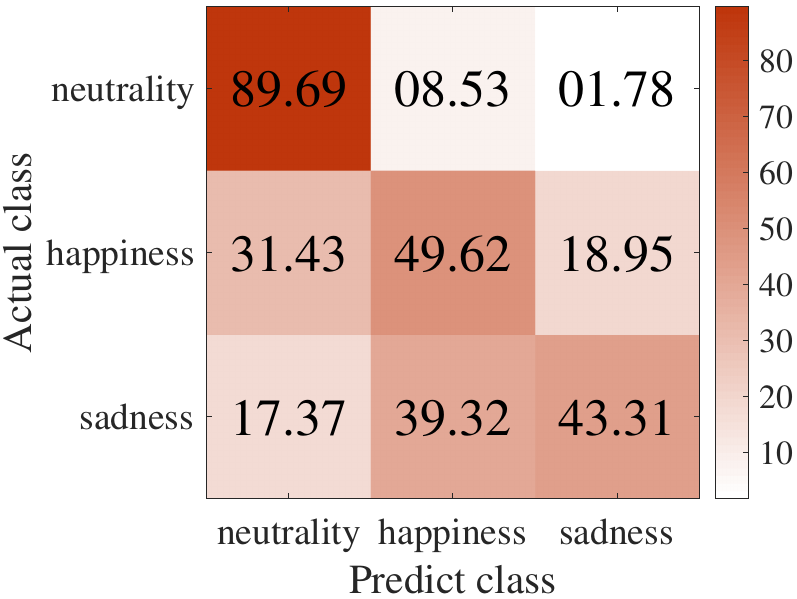}}
		\quad
		
		\centering
		\subfigure[\tiny SEED$^3$ $\rightarrow$ SEED-IV$^3$]{\includegraphics[width=0.32\linewidth]{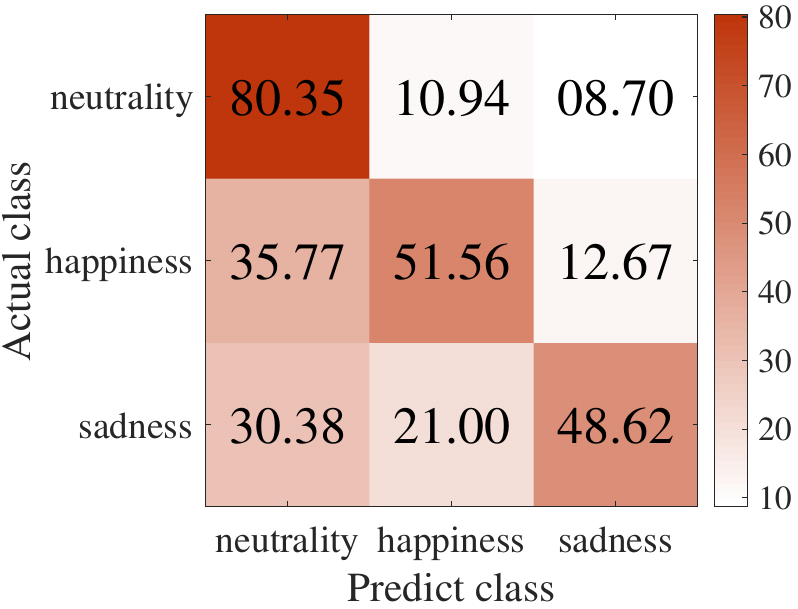}}
		\subfigure[\tiny SEED-IV$^3$ $\rightarrow$ MPED$^3$]{\includegraphics[width=0.32\linewidth]{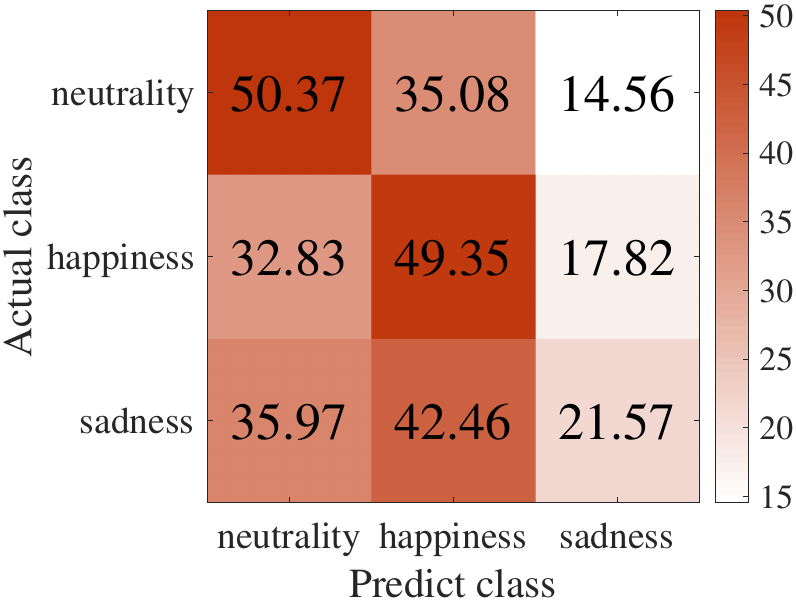}}
		\subfigure[\tiny SEED$^3$ $\rightarrow$ MPED$^3$]{\includegraphics[width=0.32\linewidth]{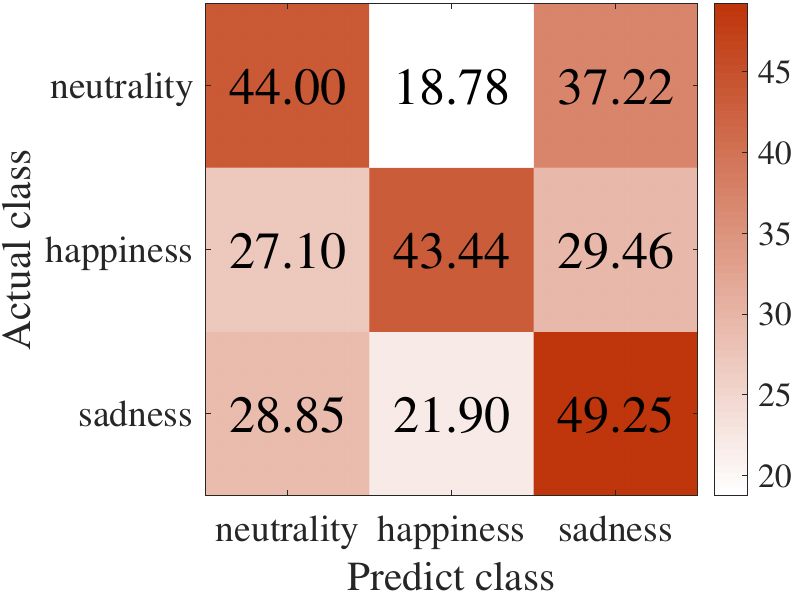}}
		\quad
		
		(1) 3-category
		
		\centering 
		\subfigure[\small MPED$^4$ $\rightarrow$ SEED-IV$^4$]{\includegraphics[width=0.4\linewidth]{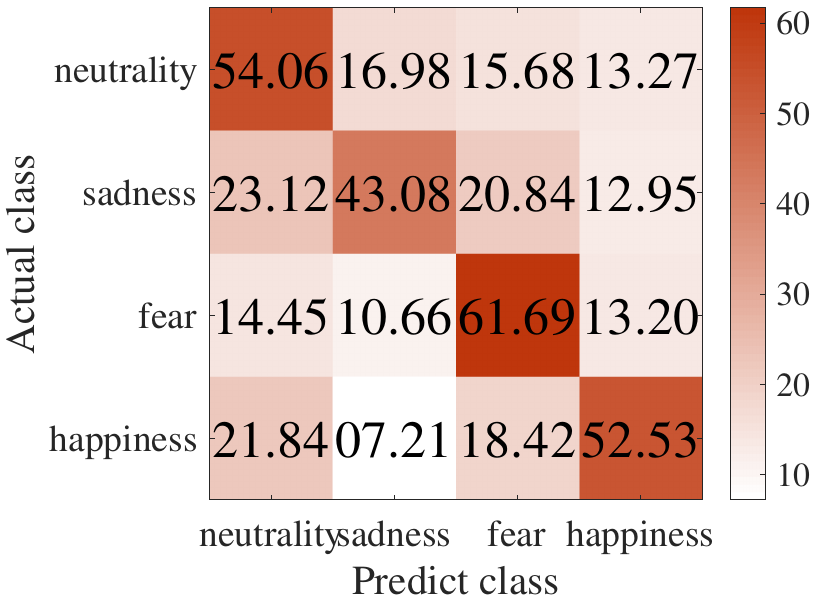}}~~~~
		\subfigure[\small SEED-IV$^4$ $\rightarrow$ MPED$^4$]{\includegraphics[width=0.4\linewidth]{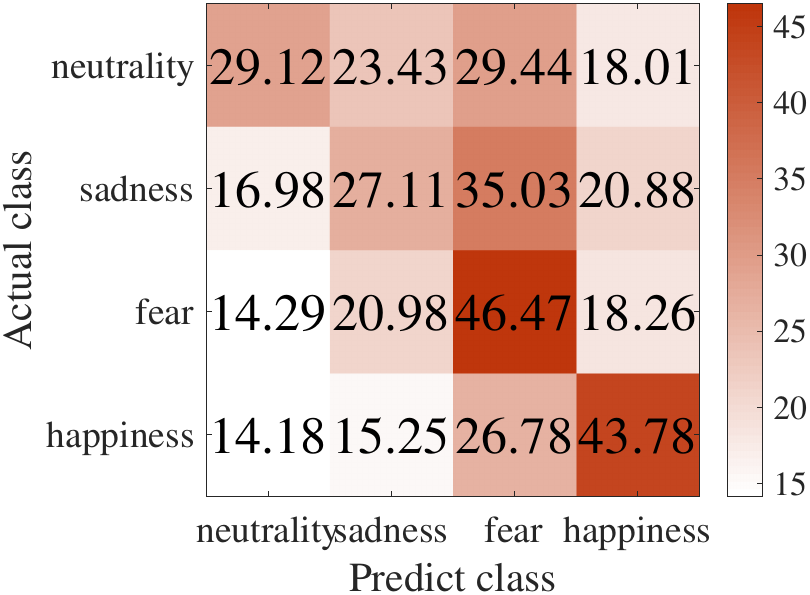}}
		\quad
		
		(2) 4-category
		
		\caption{\label{Confusion matrix}Confusion matrices of E$^2$STN results on cross-dataset experiments.}
	\end{figure}
	
	Table~\ref{Table: cross-dataset} showcases the superior performance of the proposed E$^2$STN model in cross-dataset EEG emotion recognition experiments, confirming the efficacy of the method in transfer and recognition tasks. Notably, E$^2$STN achieves the highest accuracy of 73.51\% in the \emph{MPED$^3$ $\rightarrow$ SEED$^3$} task, surpassing compared advanced algorithms significantly. In comparison with the domain adaptation method TANN, E$^2$STN demonstrates a notable accuracy improvement of 09.28\% (73.51\% vs 64.23\%) in the 3-category classification of the \emph{MPED$^3$ $\rightarrow$ SEED$^3$} task. Additionally, in the \emph{SEED-IV$^3$ $\rightarrow$ SEED$^3$} task, E$^2$STN achieves a 02.06\% (60.51\% vs 58.45\%) enhancement compared to the advanced method PGCN.
	
	In \emph{MPED$^3$ $\rightarrow$ SEED-IV$^3$} and \emph{SEED$^3$ $\rightarrow$ SEED-IV$^3$} experiments (4th and 5th columns of Table~\ref{Table: cross-dataset}), the recognition performance of E$^2$STN experiences a decline, potentially due to reduced emotional feature discrimination when the SEED-IV dataset captures finer emotion nuances. Meanwhile, the similar classification performance in these tasks (62.32\% vs. 61.24\%) underscores the effectiveness of the proposed method in eliminating the domain shift problem. Regarding the MPED dataset, which contains more emotion categories and exhibits less discrimination between emotions, resulting in a further decline in model performance (6th and 7th columns of Table~\ref{Table: cross-dataset}), E$^2$STN still outperforms other advanced methods. To validate the confidence of our experimental results, we perform the t-test statistical analysis~\cite{hanusz2016shapiro} on each reproduced accuracy result. The Shapro-Wilk test (S-W test)~\cite{semenick1990tests} is initially conducted to eliminate accuracy data that does not follow the normal distribution hypothesis. The results show that our proposed E$^2$STN exhibits significantly better ($ p < 0.05 $) performance in each cross-dataset task. This statistical analysis indicates that our proposed method effectively reduces inter-domain differences among different datasets, achieving efficient cross-dataset EEG emotion recognition.
	
	To explore which emotion is more easily recognized by the proposed model, we present confusion matrices based on the results of E$^2$STN, depicted in Fig.~\ref{Confusion matrix} (1). Several observations can be made from these matrices. Except for the \emph{SEED$^3$ $\rightarrow$ MPED$^3$} experiment (Fig.~\ref{Confusion matrix} (f)), the recognition accuracy of 'happiness' emotion is consistently higher than that of 'sadness', with an average difference of 24.56\%. This suggests that 'happiness' is more distinguishable than 'sadness' across different datasets, indicating that 'happiness' emotion is more universally induced. Furthermore, compared with 'happiness', the average accuracy of 'sadness' is 40.38\%. The lower recognition accuracy for 'sadness' is attributed to its tendency to be mistaken for 'neutrality', especially in Fig.~\ref{Confusion matrix}(a), (b), (d), and (f). This phenomenon may be due to the weak stimulation of "sadness" in these experiments. In the \emph{MPED$^3$ $\rightarrow$ SEED-IV$^3$} and \emph{SEED$^3$ $\rightarrow$ SEED-IV$^3$} experiments, similar observations between Fig.~\ref{Confusion matrix}(c) and (d) are made when SEED-IV is the target dataset. The recognition accuracies of E$^2$STN for the three emotions follow this order: 'neutrality' \textgreater 'happiness' \textgreater 'sadness', suggesting that increased emotional categories in the SEED-IV dataset result in more subtle emotional changes and increased difficulty in recognition. For the \emph{SEED$^3$ $\rightarrow$ MPED$^3$} experiment in Fig.~\ref{Confusion matrix}(f), the recognition accuracy for 'sadness' is highest, contrary to other experiment results. The greater difference in emotional categories between the SEED and MPED datasets may contribute to the more pronounced recognition of 'sadness' emotions.
	
	\subsubsection{4-category cross-dataset EEG emotion recognition}
	\label{4-category Cross-dataset EEG emotion recognition}
	
	To assess the effectiveness of E$^2$STN across a broader range of emotional categories, we conduct additional 4-category cross-dataset EEG emotion recognition experiments. In parallel, we performed comparative experiments with the same advanced methods. The results of these experiments are detailed in Table~\ref{Table: 4-categories cross-dataset}.
	
	\begin{table}[h]
		\centering
		\fontsize{10}{14}\selectfont    %{字体尺寸}{行距}
		\caption{4-category cross-dataset classification performance for EEG emotion recognition on SEED-IV and MPED.}
		\resizebox{0.48\textwidth}{17mm}
		{		
			\begin{tabular}{c|cc}
				\hline
				\hline
				\multirow{2}{*}{Method} &\multicolumn{2}{c}{\textbf{ACC / STD (\%)}} \cr
				\cline{2-3}
				& \textbf{MPED$^4$ $\rightarrow$ SEED-IV$^4$} &\textbf{SEED-IV$^4$ $\rightarrow$ MPED$^4$} \cr
				\cline{1-3}		
				SVM~\cite{suykens1999least} &~24.62/05.66 &~24.99/0.05 \cr
				BiDANN~\cite{li2018novel} &~48.56/07.73 &~32.21/06.77 \cr
				A-LSTM~\cite{song2019mped} &~35.80/06.13 &~34.07/04.55 \cr
				IAG~\cite{2020Instance} &~49.30/05.85 &~33.92/04.94 \cr				
				TANN~\cite{LI202192} &~49.40/07.33 &~33.73/01.95 \cr
				GECNN~\cite{9448460} &~50.86/08.30 &~33.13/02.65 \cr
				{PGCN~\cite{ZHOU2023126262}} &~50.32/09.04 &~36.51/07.22 \cr
				{E$^2$STN} &~\textbf{53.75/06.82} &~\textbf{36.78/04.79} \cr
				\hline
				\hline
			\end{tabular}\vspace{0cm}
		}
		\label{Table: 4-categories cross-dataset}
	\end{table}

	\begin{table*}[h]
		\centering
		\fontsize{8}{11}\selectfont    %{字体尺寸}{行距}
		\caption{Ablation experiments for 3-category cross-dataset EEG emotion recognition.}
		\resizebox{\textwidth}{9mm}
		{		
			\begin{tabular}{c|c|c|c|c|c|c}
				\hline
				\hline
				\multirow{2}{*}{Method}  &\multicolumn{6}{c}{\textbf{ACC / STD (\%)}} \cr
				
				\cline{2-7}
				&{MPED$^3$ $\rightarrow$ SEED$^3$} &{SEED-IV$^3$ $\rightarrow$ SEED$^3$} &{MPED$^3$ $\rightarrow$ SEED-IV$^3$} &{SEED$^3$ $\rightarrow$ SEED-IV$^3$} &{SEED-IV$^3$ $\rightarrow$ MPED$^3$} &{SEED$^3$ $\rightarrow$ MPED$^3$} \cr
				
				\hline	
				{E$^2$STN} &~\textbf{73.51/07.23} &~\textbf{60.51/05.41} &~\textbf{62.32/06.60} &~\textbf{61.24/15.14} &~\textbf{40.43/04.49} &~\textbf{45.56/04.78} \cr
				{E$^2$STN-t} &~65.12/08.99 &~53.70/08.53 &~55.58/08.39 &~58.89/14.35 &~38.28/04.97 &~37.85/03.71 \cr
				\hline
				\hline
			\end{tabular}
		}
		\label{Table: E$^2$STN-t}
	\end{table*}
	
	\begin{table}[h]
		\centering
		\fontsize{8}{11}\selectfont    %{字体尺寸}{行距}
		\caption{Ablation experiments for 4-category cross-dataset EEG emotion recognition.}
		{		
			\begin{tabular}{c|cc}
				\hline
				\hline
				\multirow{2}{*}{Method} &\multicolumn{2}{c}{\textbf{ACC / STD (\%)}} \cr
				\cline{2-3}
				& \textbf{MPED$^4$ $\rightarrow$ SEED-IV$^4$} &\textbf{SEED-IV$^4$ $\rightarrow$ MPED$^4$} \cr
				\cline{1-3}
				{E$^2$STN} &~\textbf{53.75/06.82} &~\textbf{36.78/04.79} \cr
				{E$^2$STN-t} &~47.39/07.13 &~34.03/05.36  \cr
				\hline
				\hline
			\end{tabular}\vspace{0cm}
		}
		\label{Table: 4-categories E$^2$STN-t}
	\end{table}
	
	In contrast to the 3-category cross-dataset EEG emotion recognition, the expansion of emotion categories results in a performance decrease for E$^2$STN. However, E$^2$STN consistently achieves the highest accuracy (53.75\% and 36.78\%) compared to other advanced methods. In the \emph{MPED$^4$ $\rightarrow$ SEED-IV$^4$} experiment, E$^2$STN outperforms the state-of-the-art method GECNN by 02.89\%. Similarly, it surpasses the state-of-the-art method PGCN by 0.27\% in the \emph{SEED-IV$^4$ $\rightarrow$ MPED$^4$} task.	Consistent with the 3-category cross-dataset EEG emotion recognition experiment, the accuracy of the MPED dataset as the target domain is slightly lower than that of the SEED-IV dataset. This may be attributed to the MPED dataset collecting more emotion categories, resulting in subtle feature differences between emotions and making transfer more challenging.
	
	For the 4-category cross-dataset EEG emotion recognition experiments, confusion matrices in Fig.~\ref{Confusion matrix} (2) reveal that 'happiness' and 'fear' emotions are easier to be recognized than 'sadness'. The 'sadness' emotion is prone to confusion with 'fear', especially in Fig.~\ref{Confusion matrix}(h), aligning with neuroscience research~\cite{KRAGEL2016444} that negative emotions (such as 'sadness' and 'fear') have quite similar Euclidean distances. Additionally, the recognition accuracy of 'neutrality' in the \emph{MPED$^4$ $\rightarrow$ SEED-IV$^4$} experiment is higher than in \emph{SEED-IV$^4$ $\rightarrow$ MPED$^4$} (54.06\% vs 29.12\%), which is reflected in the overall recognition results (53.75\% vs 36.78\% in Table~\ref{Table: 4-categories cross-dataset}).
	
	\subsection{Discussion}
	\label{Discussion}
		
	\subsubsection{Effect of the transfer module}
	To validate the effectiveness of the proposed transfer module, we modify the E$^2$STN framework, retaining only the discriminative prediction module, denoted as E$^2$STN-t. E$^2$STN-t follows the same experimental protocols as E$^2$STN but is trained solely on labeled source domain samples rather than source domain and stylized EEG samples. The experiment results are presented in Table~\ref{Table: E$^2$STN-t} and~\ref{Table: 4-categories E$^2$STN-t}. Compared with E$^2$STN-t, E$^2$STN has a substantial improvement in the performance of 3- and 4-category cross-dataset EEG emotion recognition experiments. In Table~\ref{Table: E$^2$STN-t}, E$^2$STN enhances the recognition accuracy by an average of 05.69\%, while in Table~\ref{Table: 4-categories E$^2$STN-t} of the 4-category cross-dataset experiments, the average increase is 04.56\%. This demonstrates that stylized emotional EEG representations effectively enhance the performance of E$^2$STN for cross-dataset EEG emotion recognition, providing further confirmation of the proposed transfer module's effectiveness.
	
	\subsubsection{Exploring the importance of emotion-related brain regions}
\begin{figure}[h]
		\centering 
		\subfigure{\includegraphics[width=0.6\linewidth]{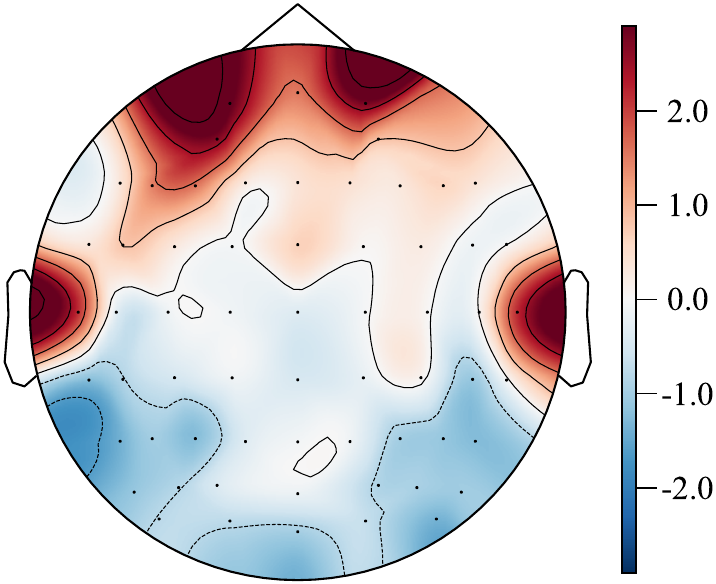}}
		\caption{\label{6}Visualization of the dynamic graph distribution in the discriminative prediction module.}
	\end{figure}

	To explore a more explicit understanding of the contribution of different brain functional regions for EEG emotion recognition, we depict the electrode activity maps in Fig.~\ref{6}. The contribution of each brain region is evident in the visualization of advanced features $ {\rm \mathbf{H_{DG}}} $, extracted by the dynamic graph convolutional layer in the discriminative prediction module. Darker red areas in the figure signify higher contributions from corresponding brain regions. The activation of the frontal and temporal lobes is prominently visible, aligning with established neuroscience research~\cite{7946165}. This observation indicates that E$^2$STN captures the most crucial emotion-related features in both source domain and stylized EEG representations, providing further evidence of the excellent performance of the proposed method for cross-dataset EEG emotion recognition.
	
	\section{Conclusion}
	\label{Sec: Conclusion}
	
	In this study, we introduce an emotional EEG style transfer network, E$^2$STN, designed to facilitate effective cross-dataset EEG emotion recognition. Three modules are constructed to accomplish the tasks of transfer, transfer evaluation, and discriminative prediction. The transfer module effectively minimizes inter-domain differences in data distribution across diverse datasets, generating stylized emotional EEG representations. The transfer evaluation module extracts multi-scale spatio-temporal features from the source domain and stylized EEG representations, constructing multi-dimensional losses to guide the emotional EEG style transfer process. The discriminative prediction module is jointly trained using both source domain and stylized EEG representations to achieve accurate prediction for cross-dataset experiments. Extensive experiments prove the effectiveness of E$^2$STN in cross-dataset EEG emotion recognition tasks. Additionally, our exploration of important brain regions related to emotion provides valuable insights into neurophysiology. For future research, we aim to delve deeper into the transfer rules of emotional EEG signals to further enhance the performance of cross-dataset EEG emotion recognition.
	
	\bibliographystyle{named}
	\bibliography{refbib.bib}
	
\end{document}